\documentclass {elsart}
\usepackage[english]{babel}
\usepackage{graphics}
\usepackage{epsfig}
\begin{document}

\begin{frontmatter}
\title{The ZEUS Micro Vertex Detector}
\author{V. Chiochia}
\collab{On behalf of the ZEUS MVD group}
\address{Deutsches Elektronen-Synchrotron DESY, 
Notkestrasse 85, D-22603 Hamburg, Germany}
\begin{abstract}
During the HERA luminosity shutdown period 2000/01 the tracking system of the 
ZEUS experiment has been upgraded with a silicon Micro Vertex Detector (MVD). 
The barrel part of the detector consists of three layers of single sided 
silicon strip detectors, while the forward section is composed of four wheels. 
In this report we shortly present the assembly procedure and in more details
the test beam results on the spatial resolution of half modules. The first
results of a cosmic ray test are presented and the radiation monitor
system is described.
\end{abstract}

\begin{keyword}
Vertex; Silicon; ZEUS
\end{keyword}

\end{frontmatter}


\section{Introduction}

\begin{figure}
	\centerline{
	\epsfig{file=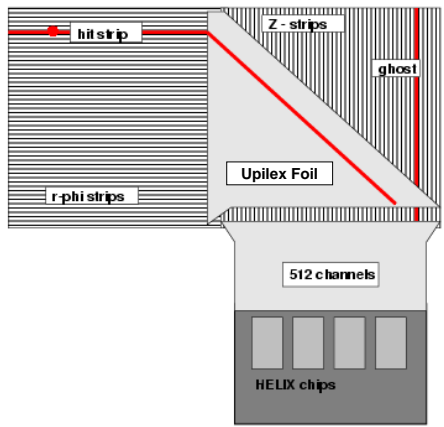,clip=,angle=0,silent=,width=6cm}
	}
	\caption{\small \it Schematic view of a half module.}
	\label{HalfModule}
\end{figure}

ZEUS is a multi purpose detector operating at the Hadron Electron
Ring Accelerator (HERA) located at the DESY laboratory in Hamburg,
Germany. At HERA 27.5 GeV electrons collide with 920 GeV  protons, 
corresponding to a center of mass energy of 318 GeV. 
The integrated luminosity recorded by the ZEUS detector 
during the first period of data taking (from 1993 to 2000) 
is about 130 pb$^{-1}$. 
A shutdown period has then followed to 
allow the luminosity upgrade of HERA \cite{Schneekloth:1998kh} 
 and the installation of new
components in ZEUS, including a silicon Micro Vertex Detector which,
besides a general improvement and extension of the track reconstruction, 
will allow the identification of final states containing long--lived
particles. \\
\begin{figure}
	\begin{center}
	\epsfig{file=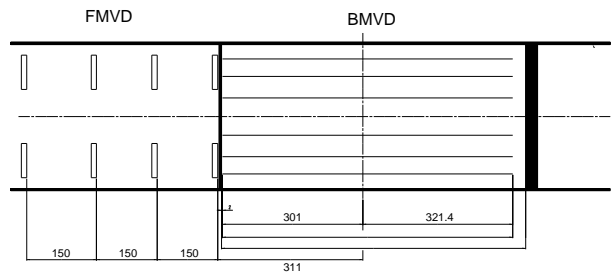,clip=,angle=0,silent=,width=8cm}
	\caption{\small \it Layout of the MVD along the beam line.}
	\end{center}
	\label{MVDlayout}
\end{figure}
Fig. \ref{MVDlayout} shows the layout of the Micro Vertex Detector (MVD) 
parallel to the beam axis. We distinguish a forward (FMVD, proton direction) and a
central (BMVD, barrel) section. The barrel section
is about 65 cm long and consists of three layers of silicon strip sensors
arranged in cylindrical planes surrounding the interaction point.
Each layer is composed by two planes of single--sided silicon strip 
detectors\footnote{produced by Hamamatsu Photonics.} (320 $\mu$m thick)
with $p^+$ strips implanted into a $n$--type bulk. The strip pitch is 20 $\mu$m but
only every 6th strip is AC coupled to an aluminium readout line. 
In the barrel section two sensors are glued together and one sensor
is electrically connected to the other via a copper trace etched on 
50 $\mu$m thick Upilex\footnote{Polyimid film produced by UBE Technologies Ltd,
Japan} foil (see Fig. \ref{HalfModule}). The connection
of the sensor assembly with the front end readout is also made via a
Upilex foil, glued at one side on the sensor and at the other side on
the hybrid. The resulting surface of 123.68 x 64.24 mm$^2$ forms one
readout cell of 512 channels and is called a {\em half module} \cite{Garfagnini:2001qv}. \\
The forward section consists of four wheels, each of them made of two
layers of 14 silicon sensors of the same type of the barrel section
but with a trapezoidal shape. In each wheel the two layers are parallel
but strips are tilted by 180$^{\circ}$/14 in opposite directions. \\
A more detailed description of the MVD layout and of the silicon sensors
can be found in Refs.
\cite{Garfagnini:1999ri,Coldewey:2000ad,Koffeman:2000tk,Koffeman:2001}.
\section{Assembly and survey}
The assembly of the barrel section proceeds as follows:
\begin{itemize}
\item Using precision 
rotational and translation tools the sensors are glued together 
to form the half module with
a precision of 10 $\mu$m. This precision is defined as the maximum
uncertainty on the distance of one reference 
mark to the one of another sensor. Two half modules (the one in Fig.
\ref{HalfModule} and its mirrored image) are mounted on top of
each other forming a {\em full module}.
\item Five full modules are fixed side by side, with 1  mm gap in between,
on a support frame and form a {\em ladder}. 
After assembly the position of the half modules are measured. Within
one ladder a precision of 5 $\mu$m is achieved. 
\item  In order to assemble the MVD and
the beam pipe together before installation in ZEUS, the detector
has been arranged in two separate half cylinders. The ladders are 
mounted in a half cylinder supported on ceramic spheres on the 
front side and by two screws on the rear. 
After assembly the ladder position is measured with a 3D survey machine
to a precision of 10 $\mu$m. A total of 30 ladders has been installed in the BMVD.
\end{itemize}
%
%
%
%
%
%
%
%
\section{Program of test and measurements before installation}
%
%
%
\subsection{Test beam measurements}

A test beam program has been set up to understand the performance
of our silicon detectors and front end electronics. 
Results on single sensors have been presented in Ref. 
\cite{Koetz:2001,Milite:2001,RedondoFernandez:2001mi}. 
Half modules of the final production chain have been exposed to an electron beam of 
2--6 GeV at the DESY II accelerator. The charged tracks are measured
by three pairs of silicon strip detectors  with 50 $\mu$m readout 
pitch and one intermediate strip, providing a position accuracy of 5.4 $\mu$m for
a 6 GeV beam. \\
The charge created by a traversing particle is collected by several 
consecutive strips, which form a {\em cluster}. 
In order to find the cluster we first look for the strip whose charge is higher then
a given threshold. The threshold was defined as 5$\sigma$, where
$\sigma  $ is the r.m.s. noise of each strip.
The reconstruction of the impact position has been performed with 
three different algorithms, i.e. (1) the Eta algorithm 
\cite{Belau:1983eh,Koetz:1985,Turchetta:1993vu}, 
(2) the Double Centroid algorithm, (3) the Head--Tail algorithm. 
\begin{enumerate}
\item {\em Eta algorithm}: is based on the assumption that most of
the charge is collected by the two readout strips closest to 
the impact position. If $Q_{left}$($Q_{right}$) is the 
charge collected on the left(right) side of the impact point, 
$P$ the readout pitch and $x_{left}$ the position of the
left strip,
the hit position $x_{Eta}$ can be reconstructed by
\begin{equation}
	x_{Eta} = P \cdot f(\eta) + x_{left}
\end{equation}
with:
\begin{equation}
	\eta = \frac{Q_{right}}{Q_{right}+Q_{left}},  \;\;\;
	f(\eta) = \frac{1}{N}\int_0^{\eta} \frac{dN}{d\eta'}\, d\eta'.
\label{eq:feta}
\end{equation}
In Eq. \ref{eq:feta} $\eta$ is calculated for each hit, while
the probability function $f(\eta)$ is evaluated using all hits of each
data sample. 
\item {\em Double Centroid algorithm (DC)}: reconstructs
the impact position using the strip with the highest collected charge
and its two neighbouring strips. Having computed the center of
gravity $C_{left}(C_{right})$ between the central strip and the left(right) 
neighbouring strip, the hit position is given by
\begin{equation}
	x_{DC} = \frac{C_{left}/ dr + C_{right}/dl}{dr+dl}, \;\;\;
	dl =\frac{Q_{left}}{Q_{right}} = 1 / dr.
\end{equation}
\item {\em Head--Tail algorithm (HT)}: 
for large incident angles, the charge generated by a particle
spreads among many strips. In this case, the charges collected by 
central strips in the cluster do not contain position information. The
Head--Tail algorithm uses the information from strips on both edges
to solve this problem. We define the head(tail) strip as a strip
which has the smallest(largest) strip number and collected 
charge higher than three times the noise level. The hit position
$x_{HT}$ is then given by
\begin{equation}
	x_{HT} = \frac{x_h + x_t}{2} + \frac{Q_t+Q_h}{2Q_{AV}} \cdot P
\end{equation}
where $x_h(x_t)$ is the position of the head(tail) strip and
$Q_h(Q_t)$ its charge. The average pulse height per strip within
the cluster is $Q_{AV}$. 
\end{enumerate}
\begin{figure}
	\begin{center}
	\epsfig{file=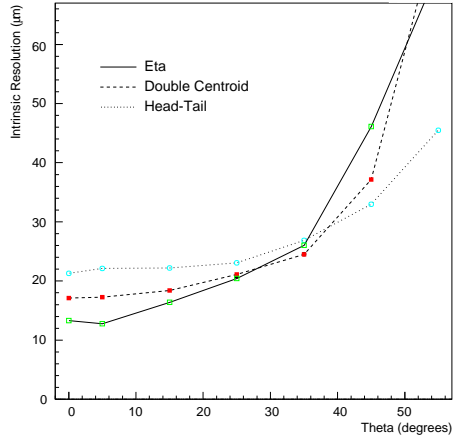,clip=,angle=0,silent=,width=6cm}
	\end{center}
	\caption{\small \it Intrinsic resolutions as a function of
	the incident angle, measured with different hit reconstruction 
	algorithms.}
	\label{resolutions}
\end{figure}
Fig. \ref{resolutions} shows the intrinsic resolution measured on
a half module as a function of the incident angle (with respect to
the normal direction) with different hit
reconstruction algorithms and a 6 GeV beam. The best achieved intrinsic 
resolution for perpendicular tracks is 13 $\mu$m with the Eta algorithm
as compared with 9 $\mu$m obtained with the single sensor \cite{Koetz:2001}.
The DC algorithm has a slightly worse resolution (17 $\mu$m) but does not
need the evaluation of the probability function Eq. \ref{eq:feta}, 
as in the case of the Eta algorithm. For incident angles greater than
40 degrees the HT algorithm gives the best performance.
\subsection{Test of half module electrical properties}
\begin{figure}[h]
\hfill
\begin{minipage}[t]{.45\textwidth}
	\begin{center}	
		\epsfig{file=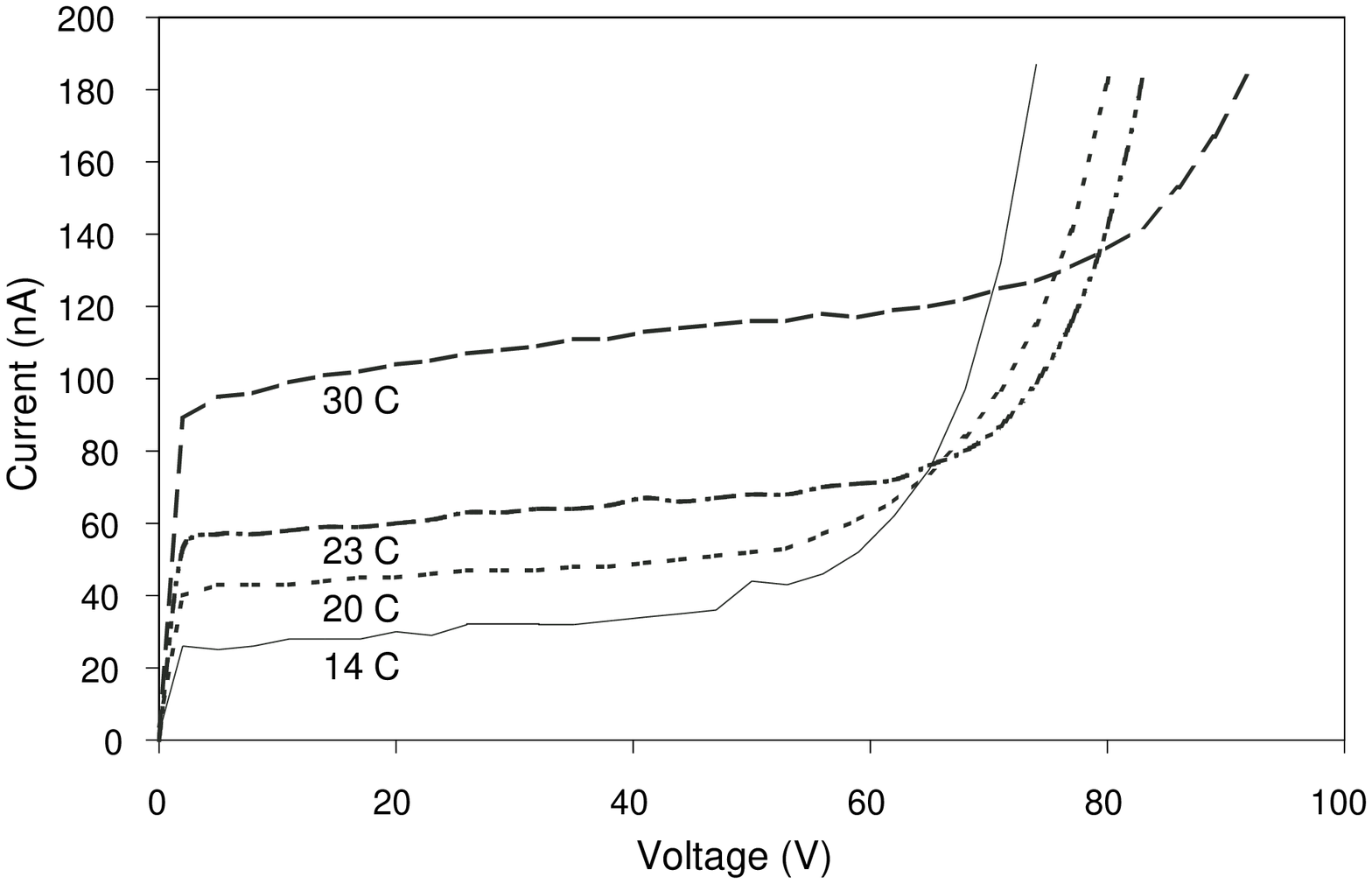,clip=,angle=0,silent=,width=6.5cm}
		\caption{\small \it I/V curves measured on an inner half module
		for five different temperatures, see text.}
		\label{iv1}
        \end{center}
\end{minipage}
\hfill
\begin{minipage}[t]{.45\textwidth}
	\begin{center}
		\epsfig{file=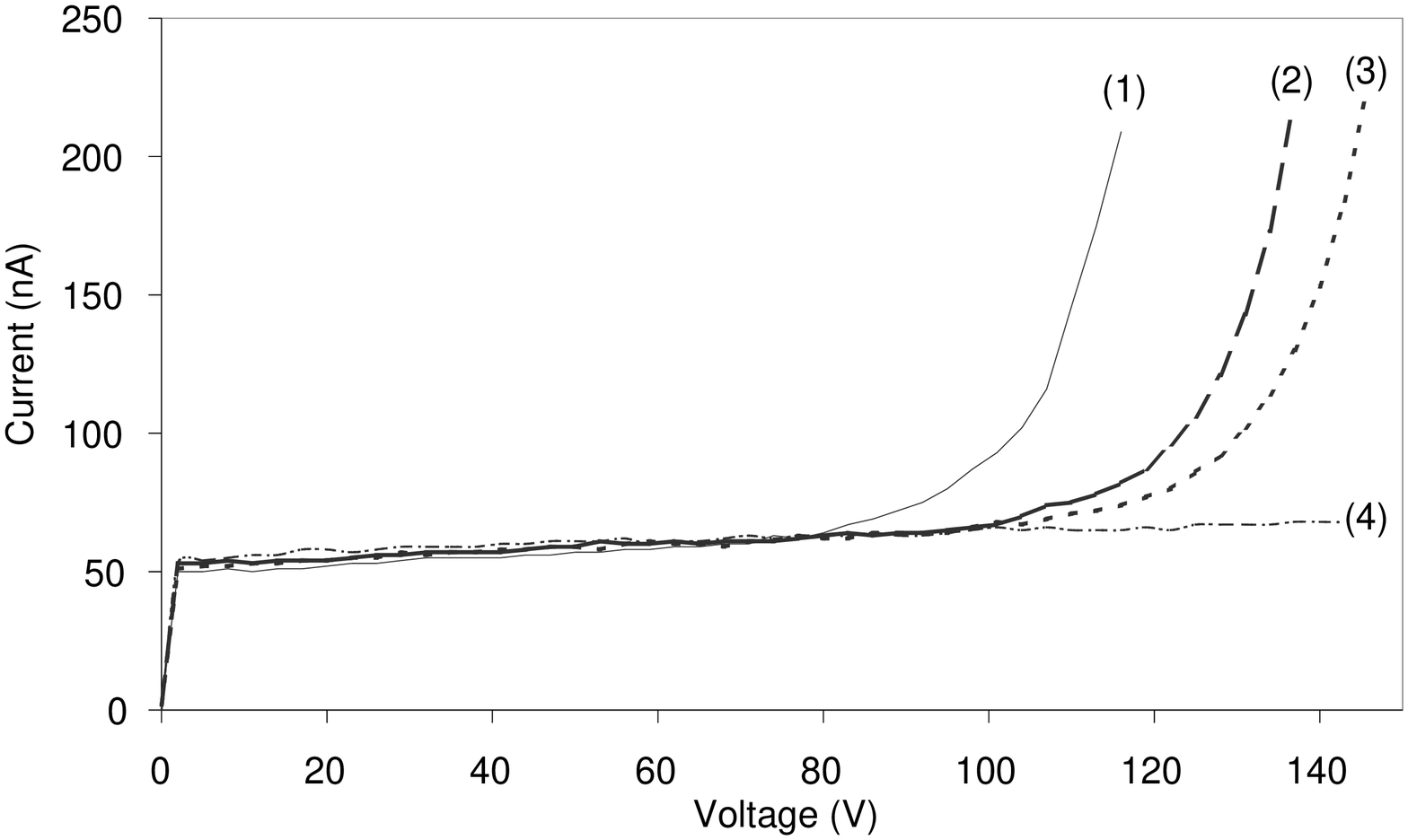,clip=,angle=0,silent=,width=6.5cm}
		\caption{\small \it I/V curves measured on an inner half module
		at 22 $^\circ$C and different relative humidity: 62.2\% (1),
		60\% (2), 55\% (3) and 31\% (4).}
		\label{humplot}
        \end{center}
\end{minipage}
\hfill
\end{figure}

A full module and two half modules (an inner and an outer one) were 
selected for a series of I/V measurements (i.e. dark current vs bias voltage)
and endurance test as a function of temperature and humidity. 
A full module was assembled in the same way as the ones used on the
ladders in the MVD. The module was placed in a box and
the temperature was monitored with a NTC resistor. 
Fig. \ref{iv1} shows the I/V curves measured on the inner half module
at different temperatures (14, 20, 23 and 30$^\circ$ C). We observe 
an increase of the breakdown voltage with the increase of temperature
as well as the expected overall increase of dark current.
During these measurements humidity was not controlled neither recorded. \\
Further measurements were performed on half modules.
The temperature in the box was regulated with a Peltier element
and humidity was monitored by a sensor Humicor S6000 
(of the type used in the ZEUS MVD). The humidity level inside the box was 
controlled by nitrogen flow. The results of these measurement are 
shown in Fig. \ref{humplot}. 
The clear dependence of the breakdown voltage on the humidity
level leads us to the conclusion that careful checking and control
of the humidity is required for the MVD.
\subsection{System test with cosmic rays}
After the assembly of the detector half cylinders together
with the beampipe the MVD has been connected to the
final readout electronics, including the final set of power
and signal cables. We summarise the tests performed in the
following points:
\begin{itemize}
	\item A fast scope test consisting in programming
	and checking the response of all front end chips;
	\item Test of data acquisition system with a random
	trigger;
	\item Data acquisition with cosmic ray trigger; 
	\item Test of slow control system.
\end{itemize}
Already from the scope test two full modules (over a 
total of 206) and a single front end hybrid have been 
found to be faulty;
no further degradation has been observed.
The system has been running continuously for four weeks in cosmic
trigger mode, recording a total of about 2.5 million events.
A first analysis of cosmic events has been performed requiring
at least two good hits in both projections in the 
modules of the outer layer. 
The hit position has been calculated using a center of gravity algorithm,
while the track trajectory has been extrapolated with a Kalman
fit. 
Input for these calculations is the geometrical position of the
detectors resulting from the design.
The distribution of residuals without any alignment correction
shows a resolution of ~70 $\mu$m in $r$--$\phi$ coordinate
and of ~80 $\mu$m in $r$--$z$ coordinate.
\section{Radiation monitoring}
The radiation monitor system (RadMon) for the MVD has the following main
tasks:
\begin{itemize}
	\item Provide continuous monitoring data and generate warning
	signals to the ZEUS shift crew in case of a moderately high 
	radiation;
	\item Provide a fast dump signal to the electron kicker of
	HERA in case the radiation is too high;
	\item Calculate the total integrated dose.
\end{itemize}
The maximum tolerable dose in 
our electronics is 3 kGy, with already degradations
in the S/N ratio after 1 kGy which can be partly recovered by tuning of the
front end parameters.
The design lifetime is 5 years of operation in the 
HERA environment, which sets the acceptable dose to 250 Gy/yr. 
The detectors chosen for the RadMon are 8 pairs of oxygen enriched 
silicon PIN diodes (1 cm$^2$ active surface, 300 $\mu$m thickness), placed
near the beampipe in the forward and rear part of the MVD. 
The two diodes forming a pair
are mounted back to back with 1 mm of lead in between
to discriminate background of charged particles from synchrotron 
radiation. \\
In addition, in the rear region six plastic tubes have been glued 
on the beam pipe shield and are used to insert TLD dosimeters, 
providing a monthly dose measurement.
\section{Conclusions}
The micro vertex detector has been succesfully assembled and
installed in the ZEUS experiment. A set of test beam measurements
on half modules shows that we have achieved an intrinsic spatial 
resolution of 13 $\mu$m for perpendicular tracks. 
The detector has been equipped with a radiation monitoring 
system and is currently participating to the commissioning 
of the HERA beams. The luminosity run will start in 2002.  
\section{Acknowledgements} 
I would like to thank all my collegues of the ZEUS MVD group,
in particular R.~Carlin, U.~Koetz, E.~Koffeman and H.~Tiecke 
for the useful discussions.
%
%

%
%
\end{document}